\documentclass[aps,twocolumn,showpacs,amsmath,amssymb,pre,superscriptaddress, floatfix]{revtex4-1}

\usepackage{graphicx}
\usepackage{dcolumn}
\usepackage{bm}
\usepackage{amssymb}
\usepackage{multirow}
\usepackage{SIunits}
\usepackage{float}
\usepackage{enumerate}
\usepackage{color}  
\usepackage{bbm}

\makeatletter
\renewcommand*\env@matrix[1][*\c@MaxMatrixCols c]{%
  \hskip -\arraycolsep
  \let\@ifnextchar\new@ifnextchar
  \array{#1}}
\makeatother

\definecolor{OutText}{gray}{0.5}

\newcommand{\mb}[1]{\mathbf{#1}}						
\newcommand{\mt}[1]{\mathrm{#1}}						
\renewcommand{\d}{\mt{d}}							

\newcommand{\Hut}[1]{\hat{#1}}

 \newcommand{\vecB} [1] {\mb{#1} }
 \renewcommand{\vec} [1] { {\vecB{#1}} }

\newcommand\myuparrow{\mathord{\uparrow}} 
\newcommand\mydownarrow{\mathord{\downarrow}} 

    \definecolor{background}{gray}{0.8}
   
\newcommand{\wdag}           { {\textcolor{white}{\dagger}} }     

\newcommand{\bra}[1]{\langle #1|}
\newcommand{\ket}[1]{|#1\rangle}
\newcommand{\ketO}[1]{#1\rangle}

\newcommand   {\<}           { \langle }
\renewcommand {\>}           { \rangle }

\newcommand{\ad}        [1] { \Hut{a} ^\dagger _{ #1 } }
\renewcommand{\a}       [1] { \Hut{a} ^\wdag   _{ #1 } }

\begin{document}

\title{Impact of Many-Body Correlations on the Dynamics of an \\ Ion-Controlled Bosonic Josephson Junction}

\author{J. M. Schurer}
    \email{jschurer@physnet.uni-hamburg.de}

\affiliation{Zentrum f\"ur Optische Quantentechnologien, Universit\"at Hamburg, Luruper Chaussee 149,
22761 Hamburg, Germany}
\affiliation{The Hamburg Centre for Ultrafast Imaging, Universit\"at Hamburg, Luruper Chaussee 149, 22761
Hamburg, Germany}

\author{R. Gerritsma}
\affiliation{Institut f\"ur Physik, Johannes Gutenberg-Universit\"at Mainz, 55099 Mainz, Germany}

\author{P. Schmelcher}
\author{A. Negretti}
\affiliation{Zentrum f\"ur Optische Quantentechnologien, Universit\"at Hamburg, Luruper Chaussee 149,
22761 Hamburg, Germany}
\affiliation{The Hamburg Centre for Ultrafast Imaging, Universit\"at Hamburg, Luruper Chaussee 149, 22761
Hamburg, Germany}

\date{\today}

\pacs{34.50.Cx, 67.85.De, 37.10.Ty, 31.15.-p}

\begin{abstract}

We investigate an atomic ensemble of interacting bosons trapped in a symmetric double well potential in 
contact with a single tightly trapped ion which has been recently proposed [R. Gerritsma \emph{et al.}, 
Phys. Rev. Lett. {\bf 109}, 080402 (2012)] as a source of entanglement between a Bose-Einstein condensate and 
an ion. Compared to the previous study, the present work aims at performing a detailed and accurate many-body 
analysis of such combined atomic quantum system by means of the ab-initio multi-configuration time-dependent 
Hartree method for bosons, which allows to take into account all correlations in the system.
The analysis elucidates the importance of quantum correlations in the bosonic ensemble and reveals that 
entanglement generation between an ion and a condensate is indeed possible, as previously predicted. 
Moreover, we provide an intuitive picture of the impact of the correlations on the out-of-equilibrium 
dynamics by employing a natural orbital analysis which we show to be indeed experimentally verifiable.
\end{abstract}

\maketitle

\section{Introduction}

In the past five years, the interest in combining ultracold atoms and ions has tremendously grown, especially 
after the first experimental attempts~\cite{Grier2009,Zipkes2010,Schmid2010} in reaching the ultra-cold 
regime 
in such hybrid atomic quantum system. Theoretical studies on the subject, however, have been carried out 
already before those experiments, for instance, for investigating related scattering 
properties~\cite{Cote2000,Idziaszek2007} 
or the formation of molecular ions in a Bose-Einstein condensate (BEC)~\cite{Cote2002}. Although not so much 
appreciated and probably not so known, this fascinating topic has already attracted the interest of 
Eugene Gross in the early sixties~\cite{Gross1962}. He was mainly interested in the estimation of the 
effective mass of a moving ion and in its impact on the ensemble of weakly repulsive bosonic particles in 
the condensate state. Gross's investigations were mostly motivated by experiments carried out at that 
time with positive and negative ions 
in liquid helium, and the goal was to develop a systematic theory starting from first principles, without 
invoking experimental data, in order to explain the large effective mass observed in corresponding low 
temperature experiments. More 
recently, however, the combination of different atomic systems like neutral atoms and ions, 
Rydberg atoms or polar molecules has open a new avenue in the research of ultracold quantum matter. Indeed, 
such hybrid quantum systems of ultracold atoms, especially atom-ion systems, 
offer a new laboratory in order to investigate ultracold chemistry and its related phenomena such as the 
formation of chemical bonds, charge exchange and transport. Furthermore, because of their 
superb controllability, combined systems of atoms and ions represent a promising platform for the study of 
condensed-matter phenomena like polarons \cite{Landau1948,Frohlich1954,Feynman1955}, the Kondo 
effect \cite{Kondo1964}, charge density waves~\cite{Bissbort2013}, and for the design of novel many-body 
quantum states. From a more genuine atomic perspective, the study of such atom-ion systems enables 
the investigation of efficient (sympathetic) cooling schemes for ions and atoms, entanglement generation and 
its propagation in the combined system (see, for instance, Ref.~\cite{Harter2014} 
for a review on the subject). 

From a many-body theory point of view, most of the current studies concerning an impurity in a condensate are 
based on approximate models like mean-field theory~\cite{Gross1962}, perturbative treatments 
\cite{Levinsen2015,Christensen2015}, or field-theoretical methods within the so-called ladder approximation \cite{Volosniev2015}. 
An interesting example for a phenomenon which can not be described in a Gross-Pitaevskii (GP) 
framework is the ionization, for instance, of a Rydberg atom in a BEC and the 
subsequent dynamics, e.g. the capturing of atoms by the ion forming mesoscopic molecular ions \cite{Cote2002}.
The new length scale induced by the long-range atom-ion interaction plays a crucial role in the static as 
well as dynamical properties of the quantum gas. Very 
recently, theoretical studies of a single impurity immersed in a BEC employing ab-initio methods like 
quantum Monte Carlo \cite{Ardila2015} or the multi-configuration time-dependent Hartree method for bosons 
(MCTDHB)~\cite{Schurer2014,Schurer2015} have been initiated. In particular, in 
Refs.~\cite{Schurer2014,Schurer2015} the static as well as the dynamical properties of a single trapped ion in 
a quasi one-dimensional Bose gas in different regimes of atom-atom interactions 
have been explored. These investigations have clearly shown the necessity to use more 
advanced methods in order to accurately describe the ground and excited states of the gas as well 
as its dynamical evolution. A simple mean-field description of such system would yield  an incomplete 
picture of the involved physics. 

\begin{figure}
\centering
 \includegraphics[width=0.45\textwidth]{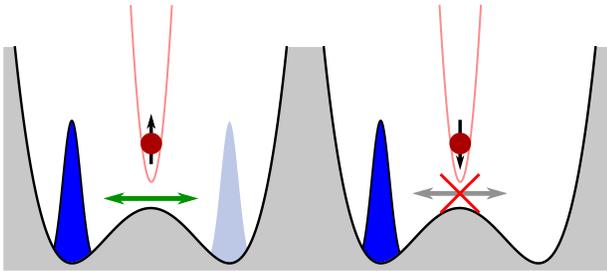}
 \caption{Setup of a bosonic Josephson junction controlled by a spatially 
localized ion in the weak link of the junction. In dependence of the internal spin state of the 
ion, the atomic ensemble can either tunnel (left panel) or is self-trapped 
(right panel).}\label{fig:setup}
\end{figure}

Motivated by these early studies and future perspectives, we analyze here in detail the many-body physics of 
an ensemble of bosonic particles in a symmetric double well potential whose tunneling is controlled by 
the internal spin state of a single tightly trapped ion such that the motion of the latter can be completely 
neglected (see Fig. \ref{fig:setup}). This setup represents essentially a controlled bosonic Josephson 
junction (BJJ). The BJJ of an atomic gas in a double well, and especially the occurrence of the macroscopic 
quantum self-trapping effect~\cite{Smerzi1997,Albiez2005}, represents a paradigmatic example of a many-body 
phenomenon, where the role played by the repulsive interatomic interactions is of paramount importance 
\cite{Zollner2006,Salgueiro2007,Sakmann2009}.
The above setup, originating from Refs.~\cite{Gerritsma2012,Joger2014}, goes one step further: by
adding a single ion in the weak link of the junction (i.e., at the barrier position, see also Fig. 
\ref{fig:setup}), the control of the atom-ion collisions can be used to 
steer the BJJ dynamics and to generate entanglement between the ion and the condensate. In particular, in 
Ref.~\cite{Gerritsma2012} it has been shown that the atom-ion 
scattering parameters can be chosen in such a way that either the tunneling regime (TR) or the self-trapping 
regime (STR) can be induced upon the internal state of the ion, and therefore 
entanglement between the two systems can be produced. Moreover, an experimentally feasible entanglement 
protocol has been proposed and validated both for a single ion and a single atom and for an ion and a 
small condensate - the latter within a two-mode Bose-Hubbard (TMBH) approach (see also 
Ref.~\cite{Milburn1997a}).
Let us also note that such a hybrid BJJ can be also thought as a basic building block of a solid-state 
quantum simulator~\cite{Bissbort2013}. 

In the present work, we are precisely interested in the question whether the control of the BJJ remains 
possible in a many-body framework (i.e., beyond mean field) with the aim of understanding if beyond mean-field 
effects are relevant in the BJJ dynamics, how they affect the temporal evolution of the TR and the STR, or 
if not to which extend the common two-mode Gross-Pitaevskii (TMGP) description is applicable. 
Moreover, we want to study the impact of the quantum correlations on the entanglement 
protocol in order to test the validity of the TMBH approximation employed in the original 
proposal~\cite{Gerritsma2012}.
We find that the control of the BJJ by the ion is indeed possible such that both dynamical regimes (i.e., TR 
and STR) can  be observed. Correlations build-up in the course of the temporal evolution and lead to 
the population of more than a single mode (i.e., orbital) rendering the TMGP approximation inadequate and 
a TMBH description questionable, especially at longer times.
Nevertheless, we show that the general entanglement protocol remains valid in its essence because the 
build-up of correlations takes place on a longer time scale than the relatively short one needed for the 
creation of the entangled state.
Furthermore, we also conclude that the description of the entanglement protocol via a TMBH
description is valid, but only if two time-dependent mode functions (i.e., orbitals) are chosen.

This work is organized as follows: In Sec.~\ref{sec:setup} we introduce the setup, thus, the model 
potential for the description of the atom-ion interaction and the many-body Hamiltonian. Here 
also the scattering properties of the two spin states of the ion are defined. Section~\ref{sec:tunneling} 
contains a detailed investigation of the dynamics of the bosonic ensemble for both internal ion states 
starting from the single-particle 
to the fully quantum correlated description. We end this section with brief considerations about
experimental strategies to measure the impact of correlations. Afterwards, the validity of the entanglement 
protocol, if quantum correlations are taken into account, is analyzed in Sec.~\ref{sec:protocol}. Our 
analysis is completed in Sec.~\ref{sec:exp} by a discussion of a possible experimental 
realization. Finally, the relevant findings and conclusions are summarized in Sec.~\ref{sec:conclusion}.

\section{Model and Theoretical Approach}\label{sec:setup}

In this section, we introduce the hybrid double well setup under consideration  and the atom-ion interaction 
which acts as the controllable ``switch'' for the dynamics of the atoms. In addition to this, we 
outline the MCTDHB method used for the simulations.

\subsection{Model System}\label{subsec:setup}

The atom-ion interaction, originating from the interplay between the charge of the ion and the induced dipole 
moment of the atom, scales in three dimensions (3D) like $-\alpha e^2/r^4$ with $r$ the distance between 
the atom and the ion, $\alpha$ the static polarizability of the atom, and $e$ the elementary charge. Let us 
note that even under strong transversal confinement, 
leading effectively to a quasi-1D motion, this interaction maintains its long-range behavior and its 1D form 
($-\alpha e^2/z^4$) is  valid up to 
some inner cutoff distance~\cite{Idziaszek2007}. Furthermore, it has been shown that the atom-ion 
scattering, 
exactly solvable by (multi-channel) quantum defect 
theory~\cite{Idziaszek2009,Gao2010,Idziaszek2011a,Gao2013,Li2014a}, can be approximated by the following 
model for the interaction potential~\cite{Schurer2014}
\begin{equation}
 \label{eq:modelPotpaper}
 V_{\mt{ion}}(z) = \mt{v}_0 e^{-\gamma z^2} - \frac{1}{z^4 + 1/\omega},
\end{equation}
which is very well-suited for our many-body investigations. Here $z$ is the atom-ion relative coordinate 
$z=z_\mt{A}-z_\mt{I}$. The parameters of the model potential $\gamma,\omega$ can be mapped onto the 
quantum defect parameters $\varphi_\mt{e}$ and $\varphi_\mt{o}$~\cite{Schurer2014} which uniquely determine 
the scattering behavior~\cite{Idziaszek2007}. The Gaussian height ($\mt{v}_0 = 3 \omega$) is chosen such that 
a node is enforced at $z=0$ which mimics the short-range behavior of the scattering solutions.
Since the ion is meant to control the BJJ, we assume it to be localized in the weak link of the 
junction (see Fig.~\ref{fig:setup}) and tightly trapped such that its motion can be neglected.
We note that the above interaction induces a length ($R^* =\sqrt{\alpha e^2 m/\hbar^2}$) and  
energy [$E^* = \hbar^2/(2m {R^*}^2)$] scale for the atoms of mass $m$ and hereafter we use them to 
rescale the Hamiltonian.  

Given this, the system of $N$ bosonic atoms in a Josephson junction with an ionic switch can be described by 
the Hamiltonian 
\begin{align}\label{eq:Hamiltonian}
  \hat{H} =& \sum_{i=1}^N \underbrace{\left[ - \frac{\partial^2}{\partial z_i^2} + V_\mt{dw}(z_i)
+  V_{\mt{ion}}(z_i) \right]}_{\hat{h}} +  g\sum_{i<j}^N  \delta(z_i-z_j) .
\end{align}
Here the Josephson junction is modeled by a double well potential of the form
\begin{equation}\label{eq:doubleWell}
 V_\mt{dw}(z) = \frac{b}{q^4} \left( z^2 - q^2 \right)^2
\end{equation}
with $b$ the barrier height  and $2q$ the  distance between the wells. This trap is designed such that 
$V_\mt{dw}(0) = b$ and $V_\mt{dw}(\pm q) = 0$. It can be approximated near the zero points by a harmonic 
potential $V_\mt{dw}^\mt{\pm q}(z) = \frac{1}{2} m \omega_q^2 (z \mp q)^2$ with $\hbar\omega_q = 4\sqrt{b}/q$.
In addition, the interaction among the atoms is of short range character and can therefore be expressed as a 
contact delta interaction of strength $g$.

For later use, we introduce a single-particle basis set $\{\phi_j(z)\}$ as eigenfunctions of the 
operator $\hat h$ 
\begin{equation}\label{eq:SPbasis}
 \hat h \phi_j(z) = \epsilon_j \phi_j(z) 
\end{equation}
with eigenenergies $\epsilon_j$. The corresponding creation and annihilation operators are defined as 
$\ad{j}$ and $\a{j}$, respectively.

A frequently employed approach to the dynamics of $N$ interacting bosons in a double well is 
the Gross-Pitaevskii equation, where the order parameter is expanded onto two single-particle mode 
functions $\ket{\mt{L}}$ and $\ket{\mt{R}}$  localized in either well of the double well 
potential~\cite{Smerzi1997}. This model 
features two main dynamical regimes: the tunneling regime in which the atom cloud oscillates between the left 
and the right well  and the self-trapping regime where most of the atomic ensemble remains trapped on one 
side of the well due to the interatomic interaction. The transition between the two regimes is characterized 
by the parameter 
$\Lambda$
\begin{equation}\label{eq:interPara}
 \Lambda = \frac{UN}{2J}
\end{equation}
which contains the on-site interaction $ U = g \int |\bra{z} \ketO{\mt{L}}|^4 \d{z}$, and the tunneling 
rate $J = \bra{\mt{L}}\hat h \ket{\mt{R}}$. A critical value $\Lambda_c$, separating the tunneling 
($\Lambda < \Lambda_c$) and the self-trapping  ($\Lambda > \Lambda_c$) regime, has been 
derived~\cite{Smerzi1997}. Starting with the initial condition that all atoms are in one well, we obtain 
$\Lambda_c = 2$. This critical value allows us to determine either a critical  interaction strength $g_c$ or a 
critical particle number $N_c$. Beyond that, it is known (see Ref.~\cite{Milburn1997a}) that in the TR the 
tunneling frequency decreases with increasing $\Lambda$, at the critical value it becomes 
zero, but it increases again for $\Lambda \gg \Lambda_c$, even though only very incomplete tunneling takes 
place.

\begin{figure*}
\centering
 \includegraphics[width=0.32\linewidth]{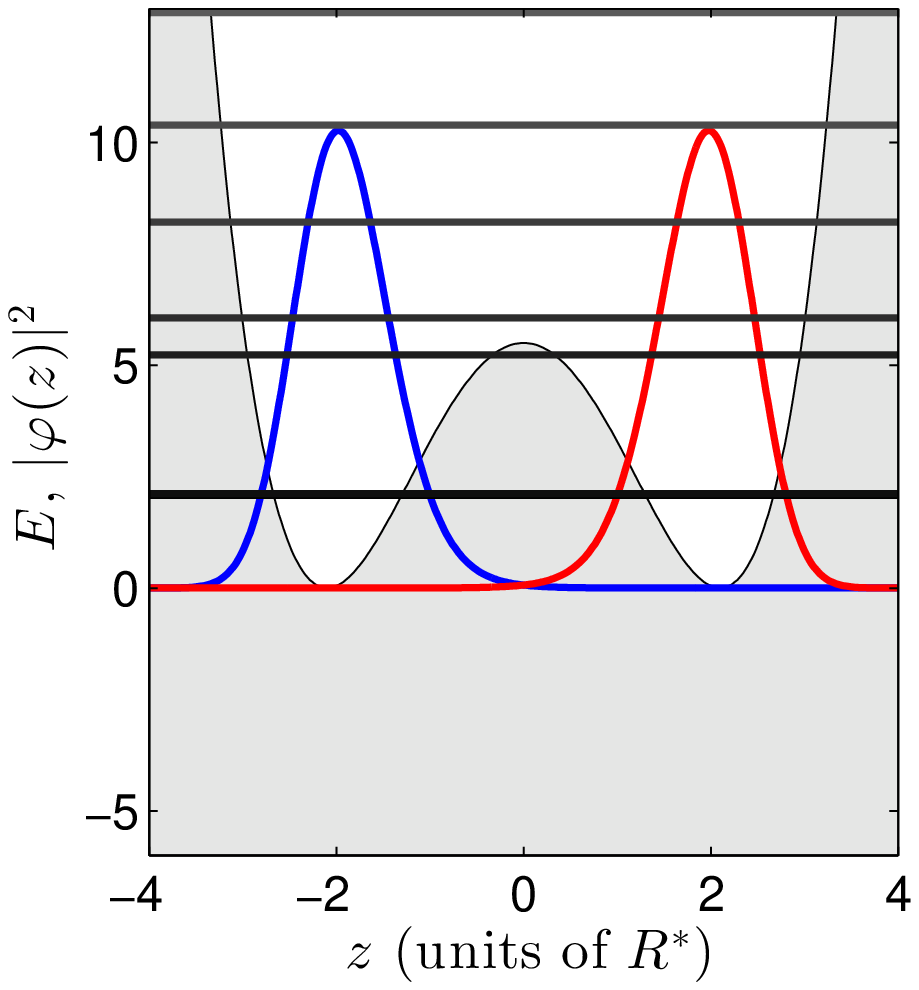}
 \includegraphics[width=0.32\linewidth]{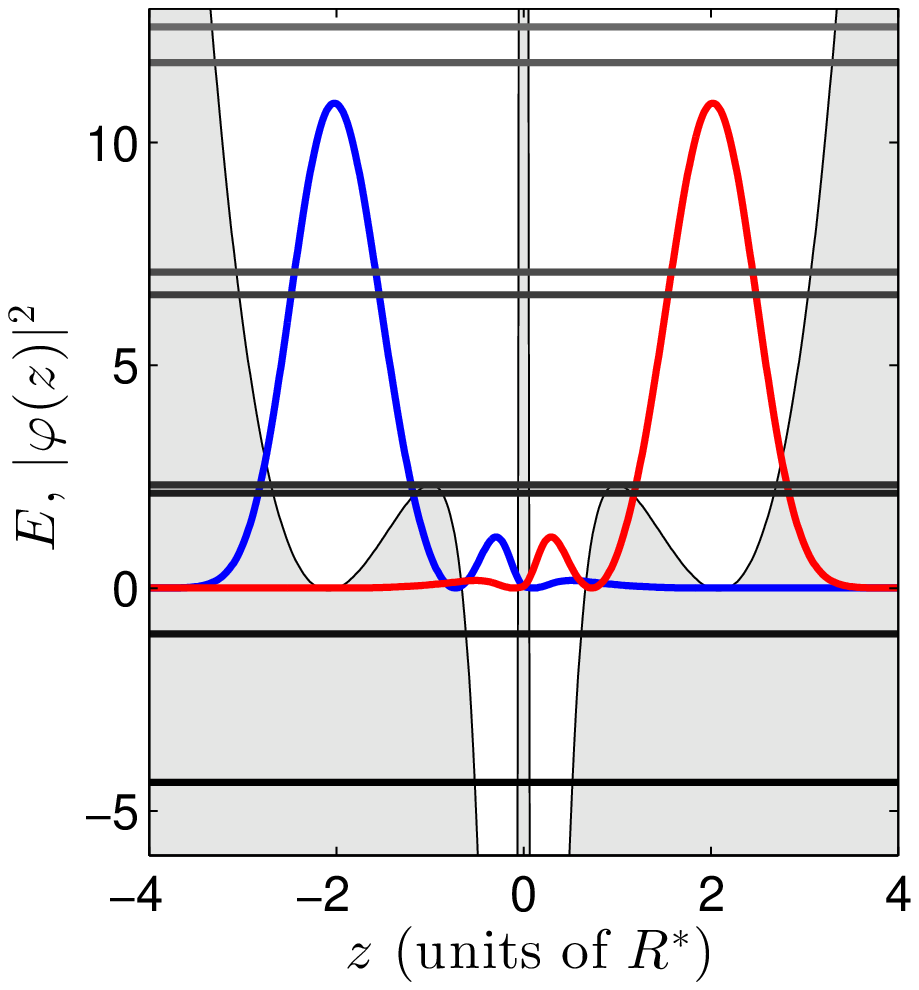}
 \includegraphics[width=0.32\linewidth]{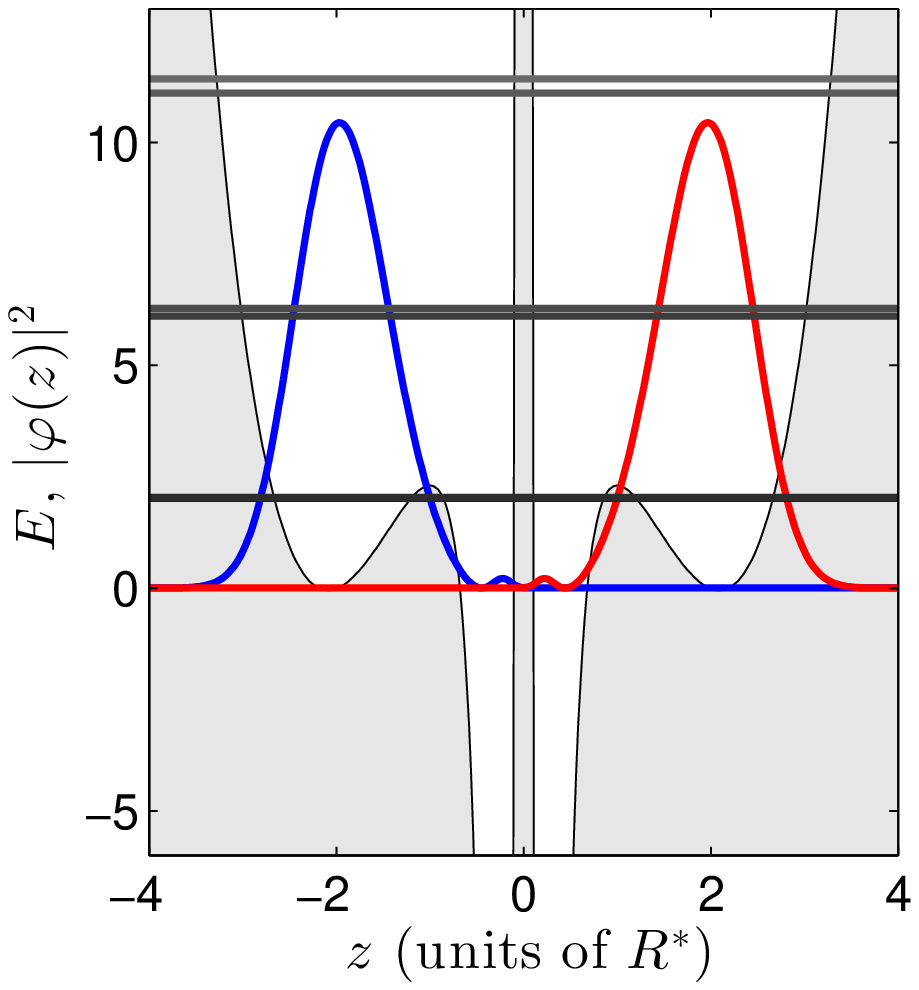}
 \caption{ Total potential (gray shaded area) together with the states $\ket{\mt{L}}$ and 
$\ket{\mt{R}}$ (blue and red line) and the single-particle eigenenergies (black horizontal lines) of the 
system without ion (left panel), with the ion in the internal state $\omega = 29\, 
(R^*)^{-4}$  and $ \gamma = 10\gamma_\mt{min}$ (middle panel), and with the ion in the internal state $\omega 
= 80\, (R^*)^{-4}$ and $ \gamma = \gamma_\mt{min}$ (right panel).
Note that in the latter case the energy of the bound states is $\epsilon_1 = 
-18.73 E^*$ and $\epsilon_2 = -15.54 E^*$.
}\label{fig:potentiallandScape}
\end{figure*}

Ignoring the ion for a moment, we choose for the entire paper the double well parameters such that the 
bosonic ensemble is in the STR with $gN=0.2 E^* R^*$ (see discussion of parameters in Sec.~\ref{sec:exp}). 
Note that the scaling of $g$ with the particle number is chosen such that $\Lambda$ is independent of 
$N$ [see Eq.~\eqref{eq:interPara}] rendering the results independent of the particle number (at least within 
a TMGP description).  
In order to allow for the ion to switch the BJJ, we further choose $\Lambda$ near, but larger than, its 
critical value $\Lambda = 3.34 >\Lambda_c$ leading to $q = 2.1 R^*$ and $b = 5.5 E^*$. With this choice 
the tunneling constant $J$ is large enough while the overlap  $\bra{\mt{L}} \ketO{\mt{R}}$ is still small.  
The corresponding potential is shown in Fig.~\ref{fig:potentiallandScape} (left panel).

Adding the ion, the objective is now to find two sets of quantum defect parameters, corresponding to two 
different short-range scattering behaviors such that in one case the atoms can tunnel through the junction and 
in the other they 
remain self-trapped. 
The quantum defect parameters can be chosen as $\varphi_e^\mt{TR} = 0.23\pi$, $\varphi_o^\mt{TR} = -0.45\pi$ 
for the TR and as $\varphi_e^\mt{STR} = 0.23\pi$, $\varphi_o^\mt{STR} = 0.3\pi$ for the STR. This choice 
corresponds to $\omega = 29\, (R^*)^{-4}$ and $\gamma = 10\gamma_\mt{min}$ in the TR leading to an 
interaction parameter $\Lambda_\mt{TR} = 0.68$ and to $\omega = 80\, (R^*)^{-4}$ and $\gamma = 
\gamma_\mt{min}$ for the STR corresponding to $\Lambda_\mt{STR}= 4.50$ with $\gamma_\mt{min} = 
4\sqrt{10\omega}$. Note, that this choice of model parameters results in two bound 
states in the atom-ion interaction potential. 

In order to understand this choice of parameters or, in other words, which property is relevant for the 
switching behavior, we look at Fig.~\ref{fig:potentiallandScape}. While 
the two double well modes $\ket{\mt{L}}$ and $\ket{\mt{R}}$ in the right panel show almost zero 
probability near the ion, those in the middle panel have a non-vanishing amplitude within the ionic 
potential. 
This can be understood by investigating the energetic position of the bound states of the atom-ion interaction 
localized in the ionic potential. In case they are near threshold, the eigenfunctions of the double well 
$\ket{\mt{L}}$ and $\ket{\mt{R}}$ obtain more and more a bound-state character which allows them to 
gain a finite amplitude in the ionic potential. We will see that in such a case the tunneling through the 
junction becomes possible.

\subsection{Theoretical Approach}

The dynamical evolution of our many-body quantum system is determined via the numerically exact ab initio 
MCTDHB method~\cite{Alon2008}. Within this method, the many-body wave function is expanded in terms of 
bosonic number states $\ket{\vec{n}(t)}$ 
\begin{equation}\label{eq:MCTDHBansatz}
  \ket{\psi} = \sum_{\vec{n}|N} A_\vec{n}(t) \ket{\vec{n}(t)}
\end{equation}
with the vector $\vec{n} = (n_1,n_2,\cdots)$ containing the occupations $n_j$ of the single-particle function 
(SPF) $\ket{\Psi_j(t)}$. The symbol $|N$ indicates that for a given $\vec{n}$ the condition $\sum_i n_i = 
N$ has to be fulfilled. Most importantly, here not only the 
expansion coefficients are time-dependent, but also the single-particle basis set $\{ \ket{\Psi_j(t)} 
\}_{j=1}^m$. Note that in Sec.~\ref{sec:tunneling} $m=5$ and in Sec.~\ref{sec:protocol} $m=4$ time-dependent 
SPFs are used assuring the convergence of all results.
By means of the Dirac-Frenkel variational principle~\cite{Dirac1930,Frenkel1934}, the variational optimal 
temporal evolution of the many-body wave function is obtained. Thereby, the coefficients $A_\vec{n}$ and the 
SPFs $\ket{\Psi_j}$ are adapted to the many-body dynamics such that even with a small number of SPFs a 
maximal overlap of the ansatz~\eqref{eq:MCTDHBansatz} to the true many-body wavefunction is guaranteed.
We refer for a detailed description of the method to Ref.~\cite{Alon2008}. Note that recently the 
method has been generalized to a multilayer (ML) structure to ML-MCTDHB allowing to treat even bosonic 
mixtures~\cite{Cao2013,Kronke2013}.

In order to analyze the high-dimensional many-body wavefunction~\eqref{eq:MCTDHBansatz}, we need to 
investigate relevant observables which allow us to understand mechanisms of the ongoing dynamics. At first, we 
use the one-body density (matrix) $\rho(z,t) = \<\hat\Psi^\dagger(z,t)\hat\Psi(z,t)\>$ ($\rho(z,z',t)= 
\<\hat\Psi^\dagger(z,t)\hat\Psi(z',t)\>$), 
defined via the field operators $\hat\Psi^\dagger(z,t)$ and $\hat\Psi(z,t)$, which gives us information 
about the spatial arrangement of the ensemble. 
Furthermore, the spectral decomposition of the reduced one-body density matrix 
$\rho(z,z',t) = \sum_j \lambda_j \Phi^*_j(z,t) \Phi_j(z',t)$ into the natural populations $\lambda_j$  and the 
natural orbitals $\Phi_j(z,t)$ provides useful information about the degree of fragmentation of the bosonic 
system~\cite{Penrose1956} and can give valuable insights into the dynamical evolution of the system.
Besides, the occupations $f_k = \< \ad{k} \a{k} \>$ and the coherences $p_{kq} = \< \ad{k} \a{q} \>$ (with 
$k\neq q$) defined with respect to the eigenfunctions $\phi_j(z)$ of the single-particle Hamiltonian $\hat h$ 
introduced in Eq.~\eqref{eq:SPbasis} will be helpful quantities to assess the excitations during the 
dynamical evolution. We remark that in the literature $\< \ad{k} \a{q} \>$ are also referred to as 
\emph{singlets} which are the matrix elements of the one-body reduced density matrix in the 
representation of $\phi_j(z)$.

\section{Tunneling Dynamics}\label{sec:tunneling}

In the following, we analyze the evolution of the tunneling dynamics of the bosonic system for the 
two sets of quantum defect parameters $\{\varphi_e^\mt{TR} = 0.23\pi$, $\varphi_o^\mt{TR} = -0.45\pi\}$ 
and $\{\varphi_e^\mt{STR} = 0.23\pi$, $\varphi_o^\mt{STR} = 0.3\pi\}$  representing two internal states of 
the ion. Thereby, we identify and analyze in detail the many-body phenomena arising in the dynamics.

Let us begin with some simple considerations about the non-interacting case. Here the singlets 
$\<\ad{k} \a{q}\>$  oscillate with the difference of the eigenfrequencies
\begin{equation}\label{eq:omega_kq}
 \hbar\omega_{kq} = \epsilon_k - \epsilon_q
\end{equation}
of the system. Note that the diagonal is therefore constant 
in time. Which of these frequencies are actually present in the dynamics depends  on the initial state.
In order to investigate tunneling dynamics, we can initialize all atoms in the state
$\ket{\mt{L}(g=0)} = \left(\ket{\phi_3} + \ket{\phi_4}\right)/\sqrt2$ which is 
localized in the left well. With this choice for the initial state we 
have $|\<\ad{3} \a{3}\>| = |\<\ad{4} \a{4}\>| = |\<\ad{3} \a{4}\>| = |\<\ad{4} \a{3}\>| = \frac{N}{2}$, while 
all other singlets vanish for all times. The dynamical evolution is here only driven by the phase of the 
coherence $\<\ad{3} \a{4}\>(t) =|p_{34}|e^{i\alpha_{34}(t)}$ which can be written as 
$\alpha_{34}(t) = \omega_{34}t$, thus, it increases linearly in time with $\omega_{34}$ since the 
$\ket{\phi_j}$ are eigenfunctions of $\hat h$ [see Eq.~\eqref{eq:SPbasis}]. It is clear that the dominant 
frequency in the process is therefore $\omega_{34}$ which is often called the \emph{tunneling frequency}. In 
the dynamics, we would observe a tunneling from the one side of the double well to the other with period 
$T_\mt{T}=2\pi/\omega_{34}$.

\begin{figure*}
\centering
 \includegraphics[width=0.45\textwidth]{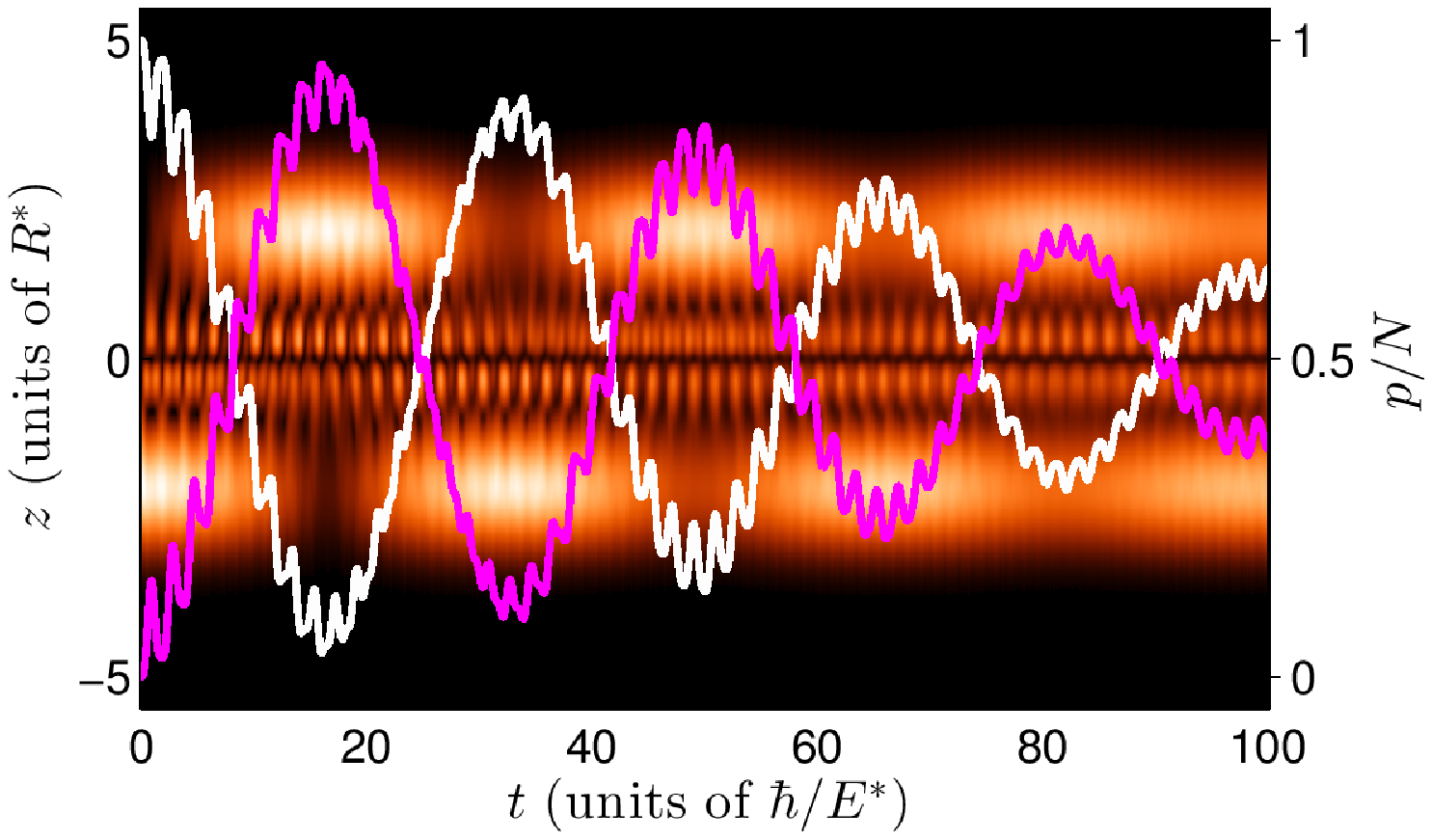}
\hspace{0.05\textwidth}
 \includegraphics[width=0.45\textwidth]{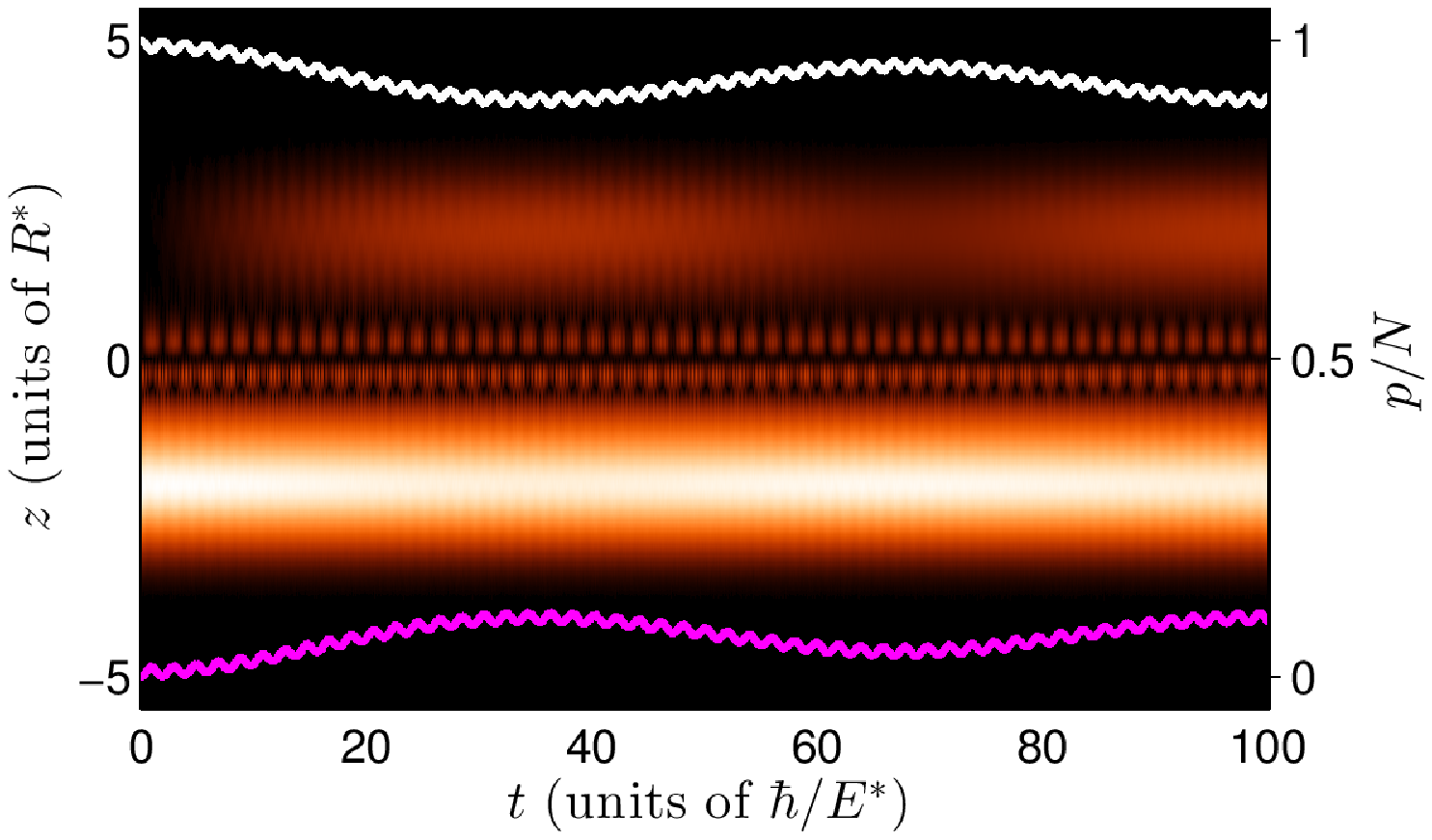}
 \caption{Temporal evolution of the reduced one-body density $\rho(z,t)$ and of $p_\mt{L}$ (white line) and 
$p_\mt{R}$ (magenta line) for $N=10$ particles for the TR (left panel) and the STR (right 
panel).}\label{fig:dynamics}
\end{figure*}

Now taking the step to the interacting many-body system defined above, the above initial 
state would be difficult to realize experimentally and therefore we start from the equilibrated ground state 
in one well. This initial state is prepared by imaginary time propagation~\cite{Kosloff1986} without the ion 
but with an additional potential blocking the right well such that the atoms 
equilibrate into the left well only. The resulting many-body wave function is used as an initial state for 
the temporal evolution with the Hamiltonian \eqref{eq:Hamiltonian} for both internal states of the ion. 
In Fig.~\ref{fig:dynamics}, we show the resulting reduced one-body density exemplarily for $N=10$ for the TR 
(left panel) and the STR (right panel). This clearly 
demonstrates that we are able to access both dynamical regimes just by changing the model potential 
parameters (i.e., the atom-ion scattering lengths).
In the dynamical process, we observe two main oscillations: a slow and a fast one. The slow oscillation 
in the TR occurs with approximatively the above discussed single-particle tunneling frequency $\hbar 
\omega_\mt{T} \approx \epsilon_4 - \epsilon_3$ ($T_\mt{T} \approx 33.3 \,\hbar/E^*$). In contrast, in the STR, 
an incomplete tunneling is observable which occurs much faster than one would expect from the single-particle 
tunneling time $T_\mt{T} \approx 225.25 \,\hbar/E^*$.
Please note that the damping of the tunneling oscillations in the TR is already a first hint that a 
TMGP description is not adequate.
The second fast oscillation mainly affects the density within the ionic potential.  Such high 
frequency oscillations have also been observed in Ref.~\cite{Sakmann2009} and attributed to the population of 
other energetically higher states. Here, however, these are attributed to the additional length and energy 
scale introduced by the atom-ion interaction, as recently reported~\cite{Schurer2015}.
Their presence is already a first indicator that even a TMBH description 
would fail to describe the above dynamics in all their details.

Let us first analyze the so-called survival 
probability $p_\mt{L}(t) = \frac{1}{N} \int_{-\infty}^0 \rho(x,t) \d{x}$ [$p_\mt{R}(t) = 
\frac{1}{N} \int_{0}^\infty \rho(x,t) \d{x}$], which allows to quantify the tunneling process. In 
Fig.~\ref{fig:dynamics}, we show $p_\mt{L}$ (white line) and $p_\mt{R}$ (magenta line) for $N=10$ particles 
for 
the TR (left panel) and the STR (right panel). We observe that in the TR $p_\mt{L}$ nearly drops down to 
zero at $t=16.5\,\hbar/E^*$ and returns back to about $90\%$ of the initial population at $t=34\,\hbar/E^*$. 
This tunneling period still nicely 
corresponds to the non-interacting tunneling period $T_\mt{T}=33.3 \,\hbar/E^* $. 
The additional superimposed oscillations are clearly visible. We will see in the following that they consist 
of more than a single frequency.
For longer times, the tunneling oscillation becomes damped 
and should reveal, at least in a TMBH description, collapse and revival behavior~\cite{Milburn1997a}.
On the other hand, in the STR (right panel), we observe that only a very incomplete tunneling occurs such 
that no population inversion is reached. The survival probability $p_\mt{L}$ is only slightly decreased to 
about $90\%$. 
Furthermore, the amplitude of the fast oscillation on top of the self-trapping dynamics is much smaller  
compared to the TR and, here, the fast oscillation contains effectively only a single frequency as it will 
become apparent below.

\begin{figure*}
\centering
 \includegraphics[width=0.45\textwidth]{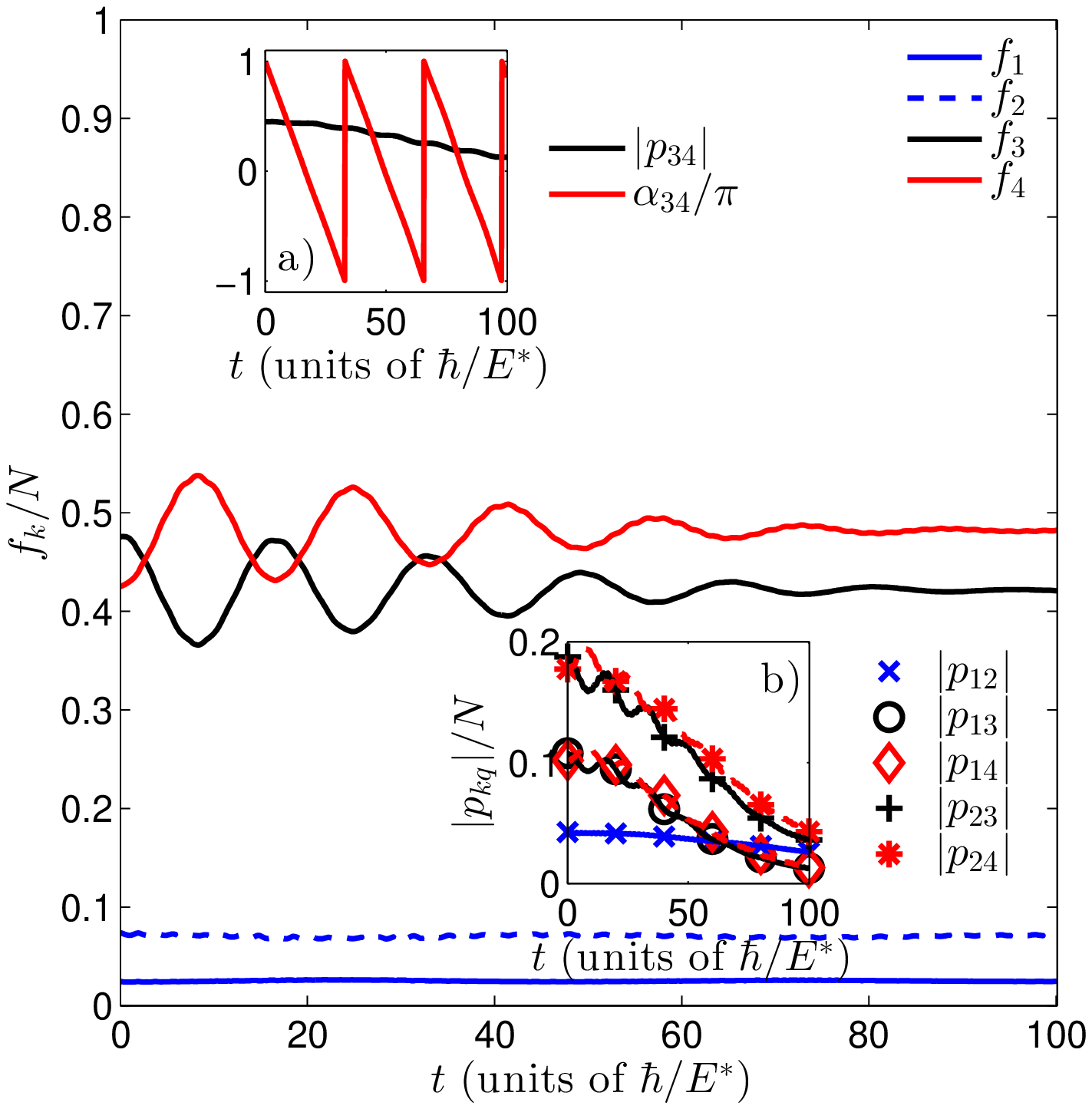}
\hspace{0.05\textwidth}
 \includegraphics[width=0.45\textwidth]{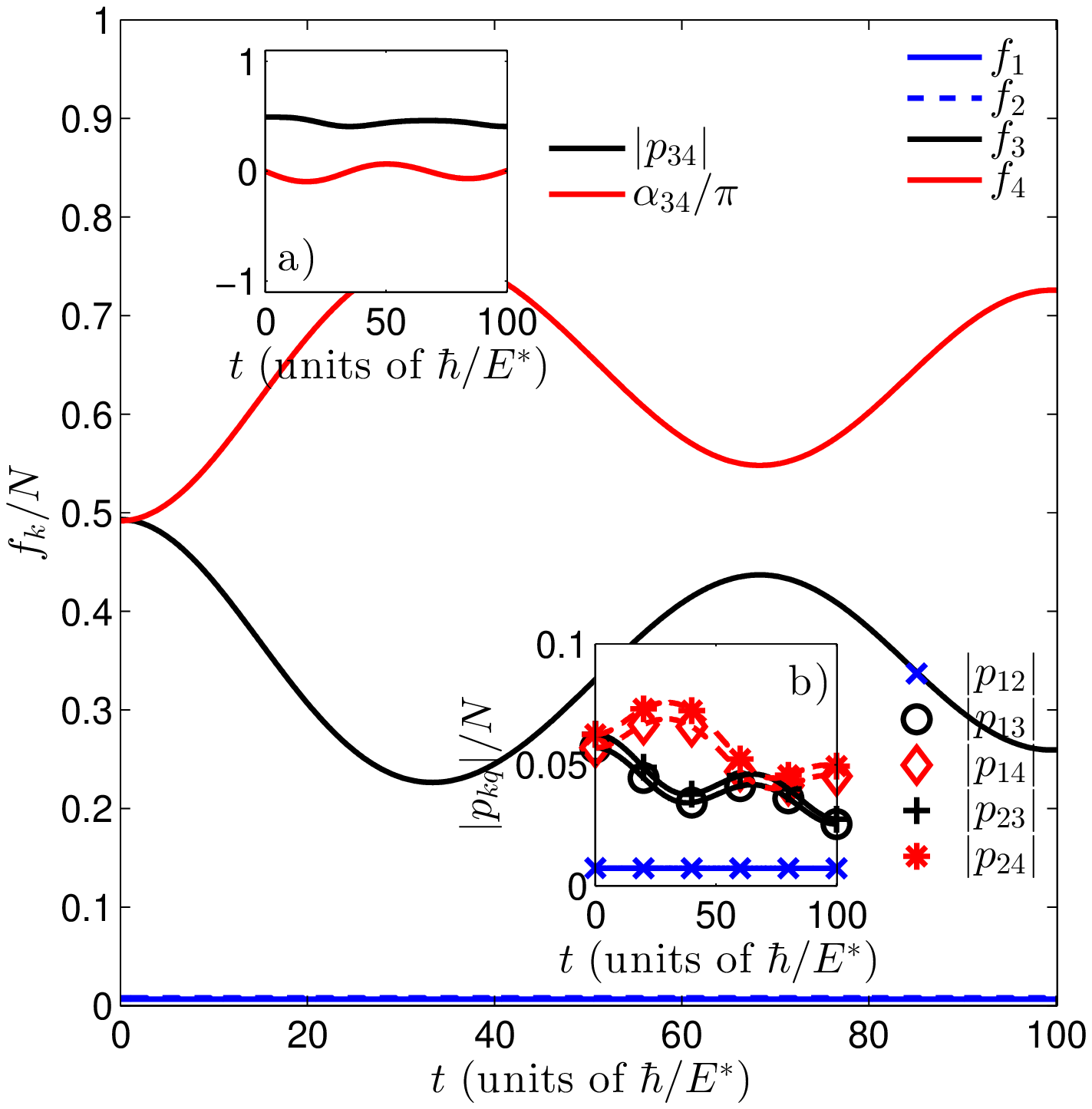}
 \caption{Temporal evolution of the dominant occupations $f_1$ (blue curve), $f_2$ (dashed blue curve), $f_3$ 
(black curve) and $f_4$  (red curve) for 
$N=10$ particles in the TR (left panel) and the STR (right panel). Upper insets: The absolute value and 
the phase of the dominant coherence $|p_{34}|$ (black curve) and $\alpha_{34}$ (dashed red curve), 
respectively. Lower insets: The absolute value of the remaining coherences $|p_{12}|,|p_{13}|,|p_ 
{14}|,|p_{23}|$, and $|p_{24}|$.}\label{fig:singlets}
\end{figure*}

In order to understand the dynamical evolution, it is helpful to investigate the contributing 
 singlets of the system. As discussed for the non-interacting case, already the initial state is here 
of importance such that different preparation schemes could change the singlet contributions. Nevertheless, 
we should keep in mind that in the many-body scenario the (absolute) value of the singlets is not necessarily 
constant in time.
In Fig.~\ref{fig:singlets}, we show the non-vanishing singlets of the two 
regimes under investigation again for the case of $N=10$ particles. It becomes clear 
that in the STR (right panel) the third and fourth single-particle eigenstates are the only 
ones which are visibly populated similarly to the single particle case. In contrast, in the TR (left 
panel), also the two bound states have, even in the initial state, a notable weight.

Moreover, we observe that, with respect to the 
non-interacting case, the occupations $f_3$ and $f_4$ show an oscillating 
behavior which is, thus, a consequence of the interactions. Using the equation of motion for the occupations 
[see Ref.~\cite{Schurer2015}] we find, under 
the assumptions that only $f_3$, $f_4$ and $p_{34}=|p_{34}|e^{i\alpha_{34}}$ ($p_{43}$) contribute, that 
quantum correlations can be neglected and by choosing real-valued single particle basis functions and  
making the approximation $|\phi_3|^2 \approx |\phi_4|^2$,  the 
evolution of the occupations is only possible as long as interactions are present, as it becomes evident from 
the following equations of motion:
\begin{align}
 \hbar \frac{\d f_3}{\d{t}} &= 2 U |p_{34}| \sin{(2\alpha_{34})}, \label{eq:f_3}\\
 \hbar \frac{\d f_4}{\d{t}} &= -2 U |p_{34}| \sin{(2\alpha_{34})}. \label{eq:f_4}
\end{align}
We see that the populations are driven by the phase of the coherences and oscillate, as long as $|p_{34}|$ is 
constant and $\alpha_{34}$ linear in time as in the non-interacting case ($\alpha_{34}(t) = \omega_{34}t$), 
as a cosine in time with frequency $2\omega_{34}$. Further, the evolutions of $f_3$ and $f_4$ possess 
opposite signs such that we can interpret Eqs. \eqref{eq:f_3} and \eqref{eq:f_4} as population transfer 
between the two states. Note that this process is not visible in the density evolution since $|\phi_3|^2 
\approx |\phi_4|^2$.
Similarly to the non-interacting case, the phase $\alpha_{34}$ in the TR increases linearly in time 
(modulo $2\pi$) with slope $\omega_{34}$ as it is visible in Fig.~\ref{fig:singlets} (upper insets) leading 
to 
clear cosine oscillations in $f_3$ and $f_4$. 
In the STR, however, where the interaction strength is effectively larger [see Eq.~\eqref{eq:interPara}] the 
phase $\alpha_{34}$ does not show a linear dependence on time anymore such that the dynamics can not be 
understood by Eqs. \eqref{eq:f_3} and \eqref{eq:f_4} only.
Therefore, we need to employ, under the same assumptions as above, the equation of motion for $p_{34}$ 
using  the equation of motion for the coherences
\begin{align}
 i \hbar \frac{\d p_{34}}{\d{t}} =& \left[ \epsilon_4 - \epsilon_3\right]p_{34} + 2 U \left(f_3-f_4\right) 
|p_{34}| \cos{(\alpha_{34})}. \label{eq:p_34}
\end{align}
The first term on the right hand side is the one which creates the linearly increasing phase $\alpha_{34}$. 
The second term comes 
into play if interactions are present and if a occupation imbalance $f_3-f_4$ between the two dominant 
modes exists. Since in the STR $|f_3 - f_4|$ can become large, the impact of this non-linear term 
which couples the equation of motion of the occupations and the one of the coherence is much more pronounced.

The fast oscillation present in Fig.~\ref{fig:dynamics} can not be explained within this two mode 
analysis. Since we have already seen that both bound states are slightly populated, especially in the TR, we 
expect these bound states to 
be the cause of the fast dynamics. Note that their occupations $f_1$ and $f_2$ are constant in time and can 
therefore not be related to the fast density oscillations visible in Fig.~\ref{fig:dynamics}. 
What remains are the coherences between 
the bound states $p_{12}$ as well as the coherences between the bound states and the two double well modes, 
that is, $p_{13},p_ {14},p_{23},p_{24}$. Their absolute values are shown in Fig.~\ref{fig:singlets} 
(lower insets). Note that their phases increase linearly in time with the non-interacting energy difference 
$\omega_{kq}$ [see Eq.~\eqref{eq:omega_kq}] as the corresponding slope (not shown).
These energy differences can be observed in Fig.~\ref{fig:potentiallandScape}. In both cases  $\omega_{13} 
\approx \omega_{14} \equiv \omega_1$ and  $\omega_{23} \approx 
\omega_{24} \equiv \omega_2$. Further, we can identify in the TR $\omega_{12} \approx \omega_{23}$ (see 
Fig.~\ref{fig:potentiallandScape} middle panel) such 
that effectively only two oscillation frequencies can be observed: $\omega_{1}$ and $\omega_{2}$. They have 
the same order of magnitude $\omega_1 \approx 2 \omega_2$.
In contrast, in the STR, the frequencies $\omega_1$ and $\omega_2$ are comparably large due to 
the large energetic separation between the bound states and the other states such that we do not resolve 
their time scale in our figures (note that the numerical simulations cover these fast time scales). 
Therefore, only $\omega_{12} \ll \omega_1 \approx \omega_2$ is visible in the shown dynamics.
Thus, the fast superimposed oscillations result in the TR from five coherences, but 
they consist only of two comparable frequencies which explain the slight beating observable in 
Fig.~\ref{fig:dynamics}. On the other hand, in the STR only a single coherence, namely $p_{12}$, is of 
importance and thus only a single frequency is visible, even though the other coherences do not vanish but 
evolve on a much faster time scale.

\begin{figure*}
\centering
 \includegraphics[width=0.45\textwidth]{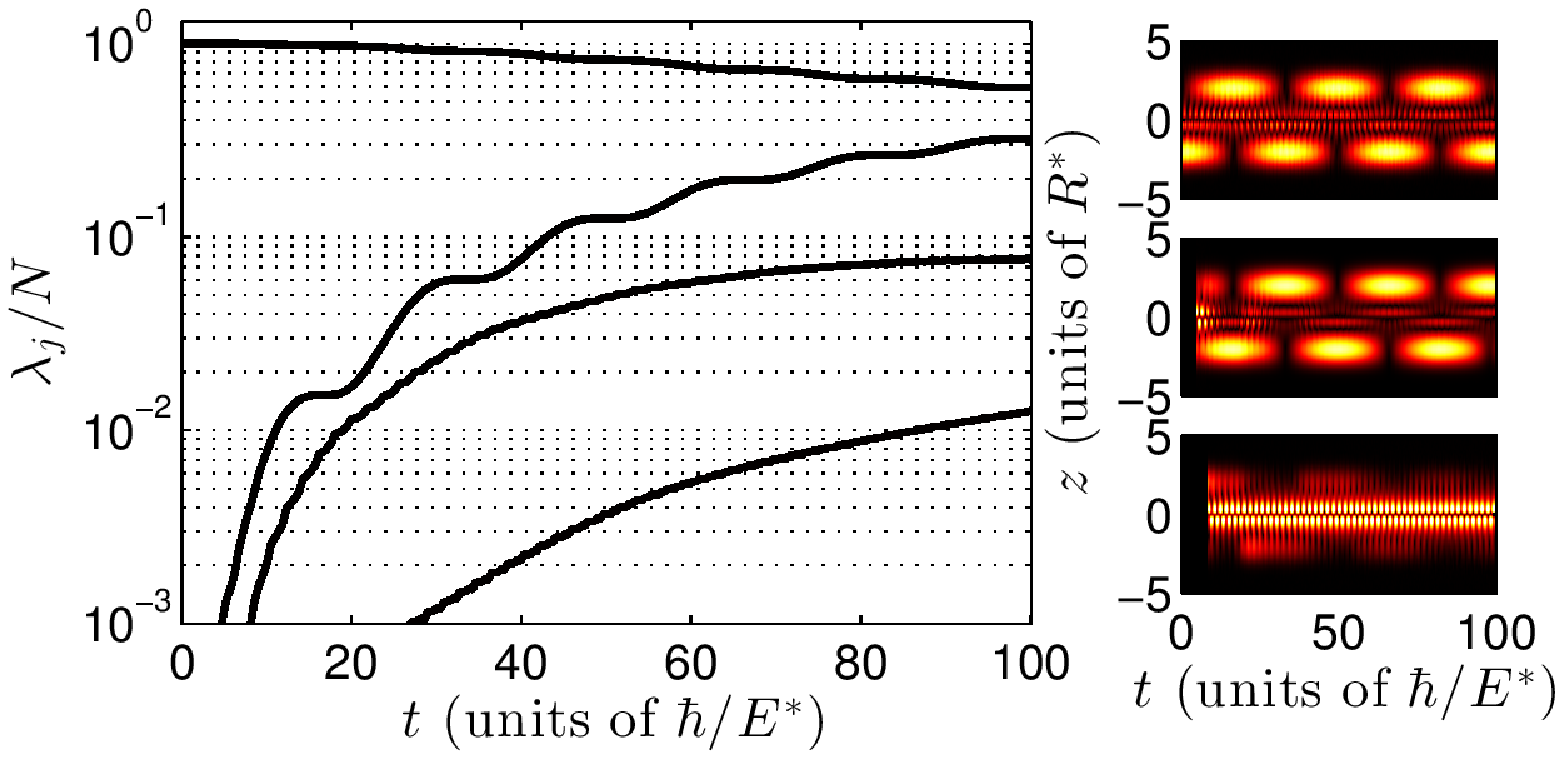}
\hspace{0.05\textwidth}
 \includegraphics[width=0.45\textwidth]{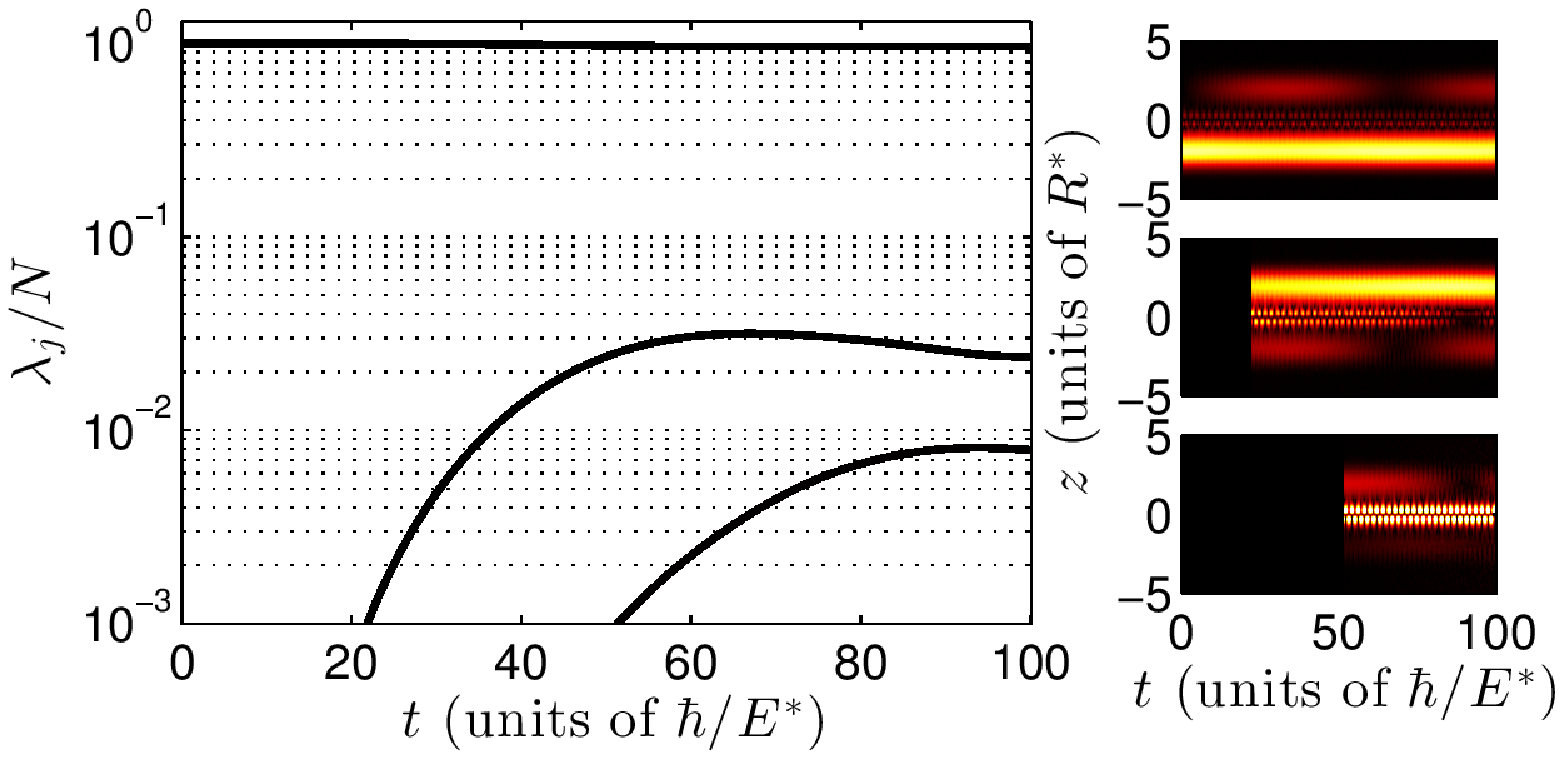}
 \caption{Temporal evolution of the natural populations (main panel; black curves) together with the absolute 
value of the associated natural orbitals $|\Phi_j(z,t)|$ [small insets; top ($j=1$), middle ($j=2$), bottom 
($j=3$)] in the TR (left panel) and the STR (right panel). Further, note 
that $\lambda_1>\lambda_2>\cdots$ by definition. The natural orbitals 
are only plotted for times in which they are considerably occupied ($\lambda_i > 
10^{-3}$).}\label{fig:natpops}
\end{figure*}

Up to now, we essentially investigated the interacting dynamics from a single particle point of view, 
namely  the singlets, which do not provide information concerning the two (and more) particle correlations. 
Nevertheless, correlations are contained in the initial state as well as build-up during the dynamics because 
of the interaction. As an indicator for beyond mean-field physics we investigate in Fig.~\ref{fig:natpops} 
the natural populations $\lambda_j$  which certify the degree of fragmentation of the 
system~\cite{Penrose1956}. Both in the tunneling and in the self-trapping regime, the initial state is nearly 
condensed ($\lambda_1 \approx 1$), since the interaction strength used is only moderate. 
Only after intermediate times, other natural orbitals become populated. For the STR a depletion of $1\%$ 
is reached in case of $N=10$ particles around $t=40\,\hbar/E^*$, whereas in the TR around $t=20\,\hbar/E^*$, 
thus much earlier than in the STR.
Moreover, the depletion in the TR becomes much more pronounced.
In both cases, we can even see the population of a third natural orbital.
In order to understand which physical processes are taking place due to the occupation of the 
additional orbitals, we investigate also in Fig.~\ref{fig:natpops} the natural orbitals 
themselves. We observe that the first natural orbital shows the behavior expected from the density of 
the atomic cloud: in the TR a clear periodic tunneling with frequency $\omega_{34}$ and in the STR a strongly 
suppressed tunneling with the majority of the atoms remaining in the left well. Since the first natural 
orbital is still the dominant one, we can think of its dynamics as corresponding to the GP solution. In this 
spirit, the second 
(and also higher) natural orbital can be understood as deviations from the GP solution induced by beyond 
mean-field correlations in the many-body wave function. Inspecting the second natural orbital, we see that it 
looks like the mirror image of the first natural orbital in both regimes.
Therefore, it corresponds to the mode which would have been dominant in case the tunneling 
dynamics had started from the other side (i.e., the right well). The fact that such a mode becomes 
populated is a clear signature that many-body effects become important and that they can influence both 
 the tunneling as well as the self-trapping behavior of the system. 
In addition to this, we observe that the third natural orbital is mainly localized within the ionic potential 
and can therefore be assigned to the above discussed contributions induced by the presence of the two bound 
states.

Let us emphasize here two important points: firstly, the presence of more than a 
single natural orbital with notable weight is a clear manifestation that a mean-field GP theory does not 
provide an accurate description of the system dynamics.
In general, however, this does not rule out a multi-orbital mean-field, as e.g. the 
best mean-field approach~\cite{Cederbaum2003}, which means that the many-body state is accurately described 
by a single permanent $\ket{\vec{n}(t)}$ [in contrast to Eq.~\eqref{eq:MCTDHBansatz}].
Secondly, the non-steady temporal evolution of the 
natural populations is a clear signature of the build-up of quantum correlations beyond a 
multi-orbital mean-field state~\cite{Kronke2015}, making the use of advanced methods such as MCTDHB 
indispensable.

In summary, despite the increasing level of many-body correlations over time in the dynamics of the bosonic 
Josephson junction, we can safely conclude that by manipulating the ion internal state a single trapped ion 
can indeed control the atomic flow through the junction.
Furthermore, we note that to the best of our knowledge the 
connection between the dynamical population of the second natural orbital and the damping of the tunneling 
dynamics has not been made previously. To this aim, we briefly discuss in the next subsection an 
experimentally viable strategy to measure the degree of fragmentation and even the approximate dynamical 
shape of the natural orbitals.

\subsubsection{Measuring the natural orbitals}

The inspection of the dynamical evolution of the natural orbitals revealed that many-body correlations in the 
dynamics lead to the build-up of a second single-particle mode which is the mirror image of the first one. 
The excitation of this process,  which seems as if the tunneling had started from the other side, effectively 
leads to an attenuation of the tunneling as well as the self-trapping behavior.
Here we aim to provide a simple strategy to experimentally verify this intuitive picture for the effect of 
the quantum correlations. 
To this end, one needs to measure the reduced one-body density matrix and diagonalize it.
In a two mode description ($\ket{\mt{L}}$,$\ket{\mt{R}}$), we can write the reduced 
one-body density matrix as 
\begin{equation}
 \hat\rho = \frac{1}{2}\left( \mathbbm{1}  + \vec{a}\cdot\vec{\sigma}\right) =  \frac{1}{2} \begin{pmatrix} 
1+a_z & a_x-ia_y\\ a_x+ia_y & 1-a_z \end{pmatrix},
\end{equation}
with the Bloch vector $\vec{a} = (a_x,a_y,a_z)$ and the vector of Pauli matrices $\vec{\sigma} = 
(\sigma_x,\sigma_y, \sigma_z)$. We note  that the reduced 
one-body density matrix does not have to be pure, thus $|\vec{a}|\leq 1$. In this model, we only need to 
measure the entries of the Bloch vector $a_i = \<\sigma_i\>$ for $i=x,y,z$.
Once those are known, the natural populations are given via diagonalization by 
\begin{equation}\label{eq:diagNatPop}
 \lambda_\pm = \frac{1}{2} \left( 1 \pm |\vec{a}| \right)
\end{equation}
with their natural orbitals
\begin{equation}\label{eq:diagNatOrb}
 \ket{\Phi_\pm} = \frac{1}{\sqrt{2|\vec{a}|(|\vec{a}\mp a_z|)}} \left[ -(a_x-i a_y)\ket{\mt{L}} + (a_z\mp 
|\vec{a}|)\ket{\mt{R}} \right].
\end{equation}

Now we only need to answer the question how and if the Bloch vector components can be measured. 
Since the left and right modes have nearly no spatial overlap, the diagonal coefficients, thus $a_z$, can be 
determined from the density $\rho(z,z)$ and define the population imbalance between the two wells. The 
off-diagonal terms, in contrast, contain the coherence of the left and the right part of the atomic cloud. 
These are accessible by time-of-flight measurements~\cite{Greiner2005} which result in, neglecting 
interactions during the expansion, the Fourier transform to $k$-space of the reduced one-body density matrix
$
    \rho(k) = \frac{1}{2\pi} \int \d{z} \d{z'}  \rho(z,z') e^{-ik(z-z')}
$ (note that $k$ and $z$ are related by the ballistic expansion condition $z(t) = \hbar t 
k/m$~\cite{Bloch2008b}).
Under the assumption that $\bra{z}\ketO{\mt{L}} \approx \bra{z-2q}\ketO{\mt{R}}$ ($2q$ is the distance 
between the wells), equivalent to $\bra{k}\ketO{\mt{L}} = 
e^{-2ikq}\bra{k}\ketO{\mt{R}}$ with $|\bra{k}\ketO{\mt{L}}| = |\bra{k}\ketO{\mt{R}}| \equiv 
|\chi(k)|$ , we obtain
\begin{equation}\label{eq:n(k)}
 \rho(k) = |\chi(k)|^2 \left[ 1 + |c|\cos{(2kq + \varphi)} \right]
\end{equation}
with $a_x +ia_y = |c| e^{i\varphi}$. Hence, by time-of-flight measurements we obtain the absolute value of 
the orbitals in $k$-space modulated with a cosine where $|c|$ is the amplitude or contrast of the modulation 
and $\varphi$ its phase. Let us draw some conclusions from these considerations: first of all, we have seen 
that the length of the Bloch vector $|\vec{a}|$ is a direct measure for the fragmentation of the system [see 
Eq.~\eqref{eq:diagNatPop}]. Second, it limits the maximal population imbalance $a_z\leq |\vec{a}|$ as well as 
the maximal contrast $|c|\leq |\vec{a}|$ in the course of the tunneling in case fragmentation is present.

Furthermore, we note that Eq.~\eqref{eq:n(k)} is not new in its form (see e.g. Ref.~\cite{Gati2007}) but 
only in its interpretation. Usually, the phase $\varphi$ is the phase between the two modes of a 
Gross-Pitaevskii 
description, whereas here it is the phase of the off-diagonal matrix element of the reduced one-body density 
matrix. Additionally, the value of $|c|$ is, in a GP description, directly connected to the diagonal elements 
of the one-body reduced density matrix, since in this case  $\hat\rho$ is constructed from a pure state.
In a fully condensed situation those two cases coincide such that Eq.~\eqref{eq:n(k)} represents a 
generalization to non-condensed (two-mode) many-body wave functions.

In summary, a measurement like the one of Refs.~\cite{Albiez2005,Schumm2005} would be sufficient to 
estimate the full reduced one-body density matrix in a two-mode approximation such that, by using 
Eqs. \eqref{eq:diagNatPop} and \eqref{eq:diagNatOrb}, the natural populations, thus the level of 
fragmentation, and natural orbitals can be obtained.
Let us note that this measurement scheme is not specific of the combined BEC-ion system investigated in this 
work, but it is quite general and can be applied to any atomic BJJ, for instance, the double well experiments 
of Refs.~\cite{Albiez2005,Schumm2005,Betz2011}. Even though the proposed measurement scheme is limited to  
two significantly populated natural orbitals, it should be sufficient for most of 
the experiments, since, typically, they are performed with a relatively large atom number and on
limited time scales such that the contribution of orbitals higher than the second one can be safely neglected.

\section{The entanglement protocol}\label{sec:protocol}

Let us investigate the possibility to generate an entangled state between the ion spin state and the 
atomic ensemble and let us denote the internal state of the ion 
leading to the TR with $\ket{\myuparrow}$ and the one leading to the STR with $\ket{\mydownarrow}$. Suppose 
now the ion to be initially prepared in a superposition of those two states $(\ket{\myuparrow} + 
\ket{\mydownarrow})$ and the atoms being prepared in the many-body ground state $\ket{\psi_\mt{L}}$ 
(or $\ket{\psi_\mt{R}}$) of the the left (or right) well.
The protocol to create a state of the form $\ket{\psi_\mt{R}}\ket{\myuparrow} + 
\ket{\psi_\mt{L}}\ket{\mydownarrow}$ was proposed in Ref.~\cite{Gerritsma2012}. There, the main 
idea was to make the distance $q$ between the two wells time-dependent. Starting at large $q$ values, 
where tunneling is suppressed, the dynamics observed above, i.e tunneling 
and self-trapping, can be enabled by reducing $q(t)$ to some fixed value $q_\mt{min}$ such that they take 
place as discussed previously. After half of a tunneling period the reverse process is performed such that the 
initial large $q$ value is restored leading finally to the desired (ideal) entangled state.

\begin{figure}
\centering
 \includegraphics[width=0.45\textwidth]{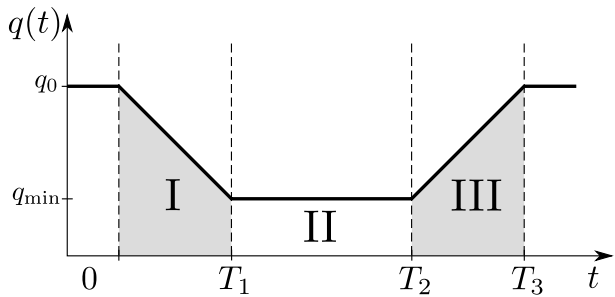}
 \caption{Sketch of the time-dependent protocol for the double well separation used to create an entangled 
state between a single ion and an atomic ensemble. Initially, the dynamics is completely suppressed at large 
$q=q_0$. During the phase $\mt{I}$, the tunneling dynamics is initiated such that it can evolve  
in the course of phase $\mt{II}$. Finally, the process is reversed in phase $\mt{III}$ by taking the 
wells far apart.}\label{fig:qt}
\end{figure}

In order to verify whether such a protocol could still work if many-body correlations are taken into account, 
we also consider here a time-dependent $q(t)$. For the sake of simplicity, we use a 
linear time-dependence, as our goal is to show the working principle of the protocol. More precisely, we 
choose the following functional form also depicted in Fig.~\ref{fig:qt}.

\begin{equation}\label{eq:qt}
 q(t) = \begin{cases} q_0  & t<0 \\ q_0 - vt & 0<t<T_1 \\ q_\mt{min} & T_1<t<T_2 \\ q_\mt{min} + v(t-T_2) 
& T_2<t<T_3 \\ q_0  & T_3<t \end{cases} 
\end{equation}

We apply the protocol outlined above to both the TR and the STR separately given the linearity of the 
Schr\"odinger equation with respect to the ion internal state.
The initial wave function in the left well is again numerically computed without the ion, blocking the 
population in the right well with a step function, and executing imaginary time propagation.
We use $q_0 = 4.5 R^*$, $q_\mt{min} = 2.1 R^*$ and $|v| = 0.1 R^*E^*/\hbar$. The time 
$T_1$ can be found via $T_1 = (q_0-q_\mt{min})/v$, while the time  $T_2$ is determined by looking at
the minimum of $p_L$ in the TR, i.e., for the ion internal state $\ket{\myuparrow}$.

\subsection{The Dynamical Evolution}\label{subsec:protocolDyn}

\begin{figure*}
\centering
 \includegraphics[width=0.45\textwidth]{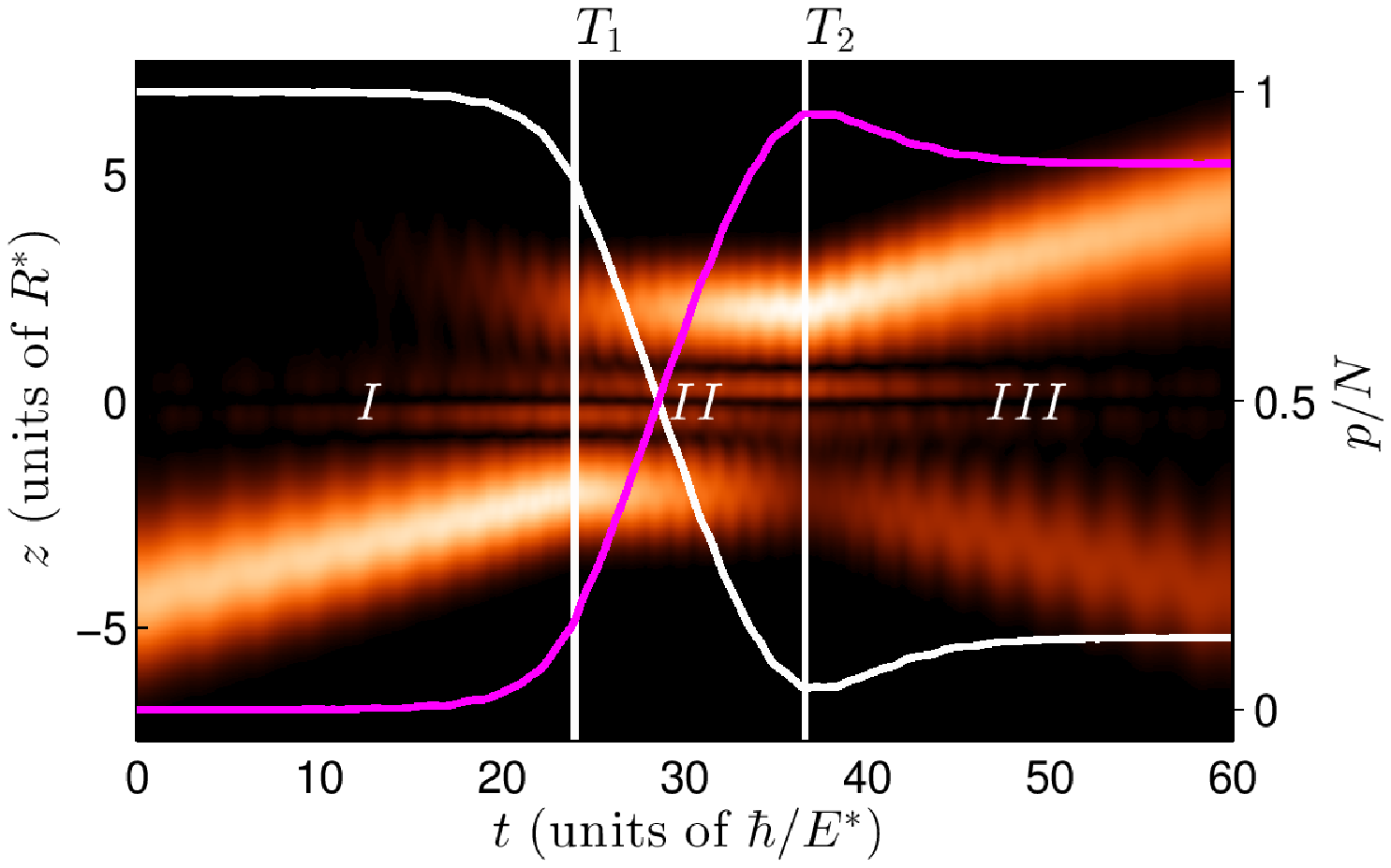}
\hspace{0.05\textwidth}
 \includegraphics[width=0.45\textwidth]{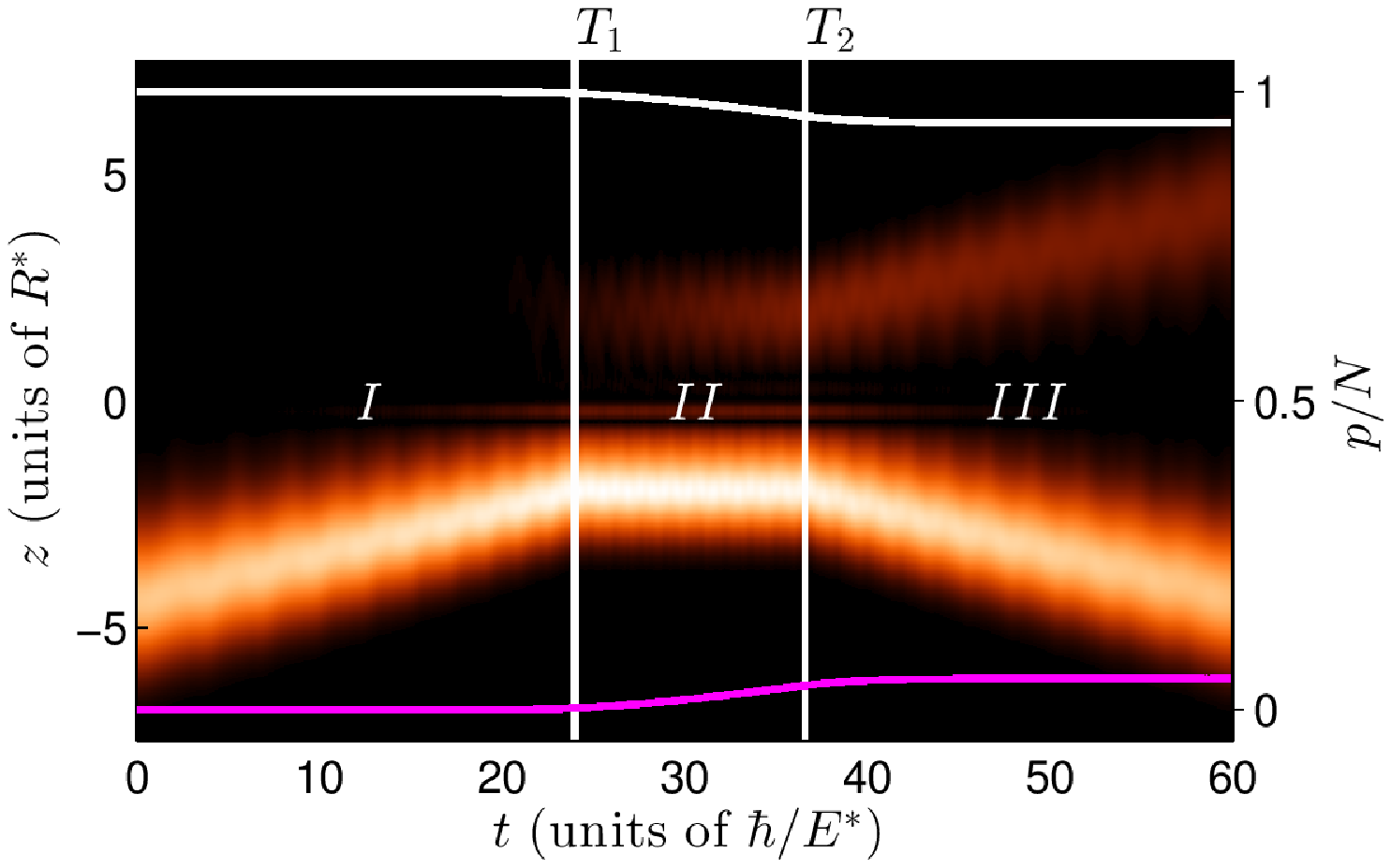}
 \caption{Temporal evolution of the reduced one-body density $\rho(z,t)$ and of $p_\mt{L}$ (white line) and 
$p_\mt{R}$ (magenta line) for $N=10$ particles using a 
time-dependent double well distance $q(t)$  for the TR (left panel) and the STR (right 
panel). The times $T_1$ and $T_2$ are indicated by white vertical lines.}\label{fig:TDdynamics}
\end{figure*}

In Fig.~\ref{fig:TDdynamics}, we show the temporal evolution of the density profile of the atomic cloud in 
the TR (left panel) and the STR (right panel) exemplary for $N=10$ using again $m=5$ SPFs. While 
for the ion internal state tuned to the TR most density is transported to the right side after the evolution, 
for  the ion in the $\ket{\mydownarrow}$ state, corresponding to the STR, most of the atoms remain 
trapped in the left well. The comparison with Fig.~\ref{fig:dynamics} shows that the dynamics occurring in 
phase $\mt{II}$ (i.e., $q(t) = q_\mt{min}$)
nearly coincides with the primary dynamics for a constant $q$. Note, however, that tunneling already 
starts in phase $\mt{I}$ in the TR. Furthermore, we highlight that in both regimes the fast 
oscillations seen in Sec.~\ref{sec:tunneling} within the ionic potential are missing here which is a 
result of the initial state preparation. Instead, small 
dipole oscillations within each well excited by the ramping procedure are visible.
Additionally,  we show the survival probabilities $p_\mt{L,R}$ of the atoms in 
the left (white line) and right (magenta line) well for the TR (left panel) and the STR (right panel) in 
Fig.~\ref{fig:TDdynamics}. We see that in the self-trapping case, the depletion of the left well population 
is only very small such that about $95\%$ of the atoms stay 
in the left well. In the TR, we nearly achieve population inversion between the two wells at $T_2$, but since 
 the dynamics cannot be instantaneously interrupted, about $12\%$ of the 
atoms are in the left well at the end of the protocol.
Although this is out of the scope of the present study, we note that higher transport efficiency can be 
attained by means of a compensating control pulse technique, that is, by optimally controlling the energy 
offset between the two wells~\cite{Brouzos2014}.

In order to understand at which level of approximation the above dynamics can be described correctly, we 
investigate the natural populations as well as the singlets of the system. Firstly, we find that in the STR 
only a single natural orbital is considerably occupied ($\lambda_1 \approx N$) meaning that the GP 
approximation is indeed fine. In contrast, in the TR, a second natural orbital becomes populated 
during the protocol up to $\lambda_2 \approx 0.05N$ such that a mean-field description is not adequate, even 
though the initial state is almost perfectly condensed.  Secondly, we checked the validity of the TMBH 
approach by the singlets 
populations $f_k$. Note that since the double well potential is now time-dependent, the 
single-particle Hamiltonian $\hat h$  and its eigenfunctions become also time-dependent [see 
Eq.~\eqref{eq:SPbasis}], and therefore the creation and the annihilation operators $\ad{j}$ and $\a{q}$, too. 
In both regimes, we find that $f_3 + f_4 \approx N$, such that only the two 
modes [$\phi_3(z,t)$ and $\phi_4(z,t)$] are sufficient for a correct theoretical description. We would like 
to emphasize, however, that only by choosing time-dependent mode functions the TMBH description of the 
protocol is valid, that is, time-independent mode functions would not lead to the correct BJJ dynamics.

Finally, let us mention that we have performed simulations of the protocol up to $N=100$ particles 
without any major qualitative difference. Only the degree of fragmentation reduces for increasing $N$ due to 
the chosen scaling of the interaction strength with the particle number (i.e. $gN = const$) such that the GP 
approximation is constantly improving for growing particle number.

\subsection{Fidelity of the Process}

Finally, we want to evaluate quantitatively the efficiency of the entanglement protocol.
The natural way to make such an assessment would be to compute the overlap integrals 
between the evolved states up to time $T_3$ with the corresponding target states. This approach turns out to 
be computationally quite expensive due to the time-dependent SPF used in our method. In order to define a 
fidelity measure that contains more information about the system state than the density, we rely on the 
so-called Uhlmann fidelity~\cite{Uhlmann1976,Jozsa1994,Nielsen2000} 
\begin{equation}\label{eq:fidelity}
 F(t;\hat\sigma,\hat\rho_\mt{G}) = \mt{Tr} \left[  \sqrt{\sqrt{\hat\rho_\mt{G}} \hat\sigma(t) 
\sqrt{\hat\rho_\mt{G}}  }  \right]
\end{equation}
Here $\hat\sigma(t)$ is the density matrix of the system at time $t$ and $\hat\rho_\mt{G}$ is the density 
matrix of the target state. In case of pure states ($\hat\sigma =\ket{\psi}\bra{\psi} $ and $\hat\rho_\mt{G} 
= \ket{\psi_\mt{G}}\bra{\psi_\mt{G}}$) it is identical to the overlap integral
$ F(t) = | \bra{\psi_\mt{G}}\ketO{\psi}|$.
In the present work we replace the density matrices $\hat\rho_G$ and $\hat\sigma$ with the corresponding  
one-body reduced density matrices. The derivation of Eq.~\eqref{eq:fidelity} using $\rho(z,z',t)$ is shown in 
App.~\ref{app:uhlman}. We note that with this choice at least the coherence, i.e. the off-diagonal 
elements of the reduced one-body density matrix, is taken into account, contrarily to a fidelity measure 
based on the atomic density (i.e., $p_\mt{L}$ and $p_\mt{R}$) only. 

\begin{figure}
\centering
 \includegraphics[width=0.45\textwidth]{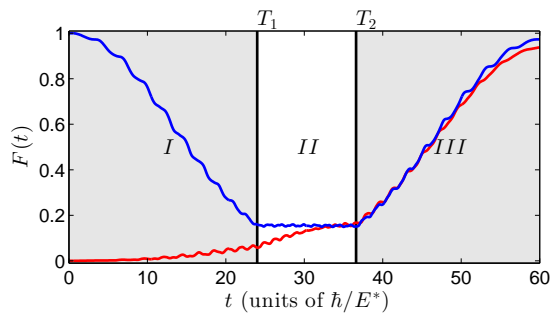}
 \caption{Temporal evolution of the Uhlmann fidelity $F(t)$. (Red curve) Fidelity to find the atomic ensemble 
in the right well in the TR. (Blue curve) Fidelity to find the atomic ensemble 
in the left well in the STR.}\label{fig:TDfidl}
\end{figure}

For the STR, we choose as the target one-body density matrix the one corresponding to the initial state, 
while for the TR the one-body density matrix corresponding to the ground state of all atoms in the right well 
is used. In Fig.~\ref{fig:TDfidl}, the fidelity in dependence of time is shown for 
the TR (red line) and the STR (blue line). We see that for the STR the initial density matrix is 
recovered with a probability of more than $F(T_3) \approx 97\%$ after the protocol. In the TR, we can prepare 
the opposite one-body density matrix with a fidelity of more than $F(T_3)\approx 93\%$.
These values could still  be improved, e.g. by optimally choosing the time $T_2$. 
Moreover, the whole ramping procedure could be fully optimized to reach a maximal fidelity by looking for an 
optimal shape of the separation $q(t)$. Nonetheless, even with the above non-optimized and simple strategy, 
the efficiency of the entanglement protocol is quite good. This shows that even without sophisticated control 
designs a very satisfactory and experimentally easy scheme can be accomplished in the laboratory with a high 
success probability.

Summarizing, we can state that when the ion is initially prepared in the state $\ket{\myuparrow} + 
\ket{\mydownarrow}$ both of the above processes would take place simultaneously such that we end up in the 
entangled state $\ket{\psi_\mt{R}}\ket{\myuparrow} + \ket{\psi_\mt{L}}\ket{\mydownarrow}$ with high fidelity.

\section{Experimental Implementation }\label{sec:exp}

The setup considered in this paper may be implemented in the laboratory by combining trapped ion technology 
with optical traps for atoms~\cite{Joger2014}. In recent years a number of experiments aimed at combining 
trapped ions and atoms have become available~\cite{Grier2009,Zipkes2010,Schmid2010, Harter2014} demonstrating 
the feasibility of this approach. Optical traps~\cite{Albiez2005,Gati2007,Levy2007} or magnetic traps derived 
from micro-structured electromagnets~\cite{ Betz2011,LeBlanc2011} may be used to create atomic double well 
potentials with trapping frequencies in the kHz regime. The ion, in turn, may be trapped in a radio frequency 
Paul trap leading to very strong confinement. An interesting alternative is to trap the ion by optical 
potentials which was recently demonstrated ~\cite{ Schneider2010,Enderlein2012}.   Employing such a 
setup may mitigate potential heating problems due to the time-dependent trapping field of the Paul trap ~\cite 
{Cetina2012, Krych2015,Joger2014}.  The internal state of the ion may be probed by fluorescence detection 
and manipulated by external electromagnetic fields, whereas the atomic density and atom number can be obtained 
by time-of-flight analysis and absorption imaging.
As a particular example we consider a $^{171}$Yb$^+$ ion interacting with a BEC of $^{87}$Rb 
atoms~\cite{Zipkes2010}. For this combination, the typical length scale of the interaction is given by $R^* = 
375.31~\,\nano\meter$ and $E^*/h=0.41\,\kilo\hertz$. Using $q=2.1 R^*$ and $b=5.5 E^*$, as before, we get a 
trapping frequency of $\omega_q=2\pi~1.84\,\kilo\hertz$ in each well. A transverse trapping frequency of 
$\omega_{\perp}\sim2\pi\,25\,\kilo\hertz$ allows us to reach the 1D regime with a 1D scattering length of 
$a_{1D}=-389.3~\,\nano\meter$~\cite{Olshanii1998}. For this choice of parameters, we obtain, using $N = 10$ 
particles, $gN \approx 0.2 E^*R^*$ as chosen in Sec.~\ref{subsec:setup}. The timescales in the manuscript 
correspond to $\lesssim 40 \,\milli\second$, for the considered atom-ion combination, such that we may expect 
coherence to be maintained in the ion’s internal state during the atomic tunneling. Heating of the ion during 
this period may be mitigated by using a large ion trap with large ion-electrode separation.

\section{Conclusion}\label{sec:conclusion}

We presented a many-body study of an atomic BJJ interacting with a single trapped ion localized 
in the center of the junction. We found that the controlled switching of the BJJ by the ion is possible 
as in the original proposal~\cite{Gerritsma2012}, namely that the TR as well as the STR can be reached for 
fixed particle number and inter-atomic interaction strength even if all correlations are taken into account. 
In the TR, we found that correlations build-up a second 
mode which decreases the population imbalance between the two wells leading to an ``equilibration'', as 
already found within a TMBH approximation~\cite{Milburn1997a}. Additionally,
a third mode  with non-zero contribution is found, which is localized within the ionic potential, leading 
to fast oscillations on top of the tunneling dynamics. Similarly, the STR dynamics is harmed by 
the presence of a second mode, even though in this case the build-up of the many-body correlations takes 
place on a longer time scale. Also here fast oscillations are superimposed which are localized in the 
ionic potential, but they are less pronounced than in the TR. Interestingly, we find that in both 
regimes the second mode arising in the dynamics is the mirror image of the first one. This shows that 
the presence of correlations triggers
 the spatial mirror physical process which one would expect to be only present if the dynamics had started 
from the other side of the double well. In order to allow for an experimental confirmation of this behavior, 
we proposed a scheme to measure the natural populations and natural orbitals.
The third mode, however, can be understood as a manifestation of the additional length and energy scale 
induced by the atom-ion interaction which 
highlights the necessity to describe the atom-ion interaction not simply by a repulsive contact interaction.
In both cases, we can conclude that a GP description would not be able to capture the dynamics 
correctly. Even a multi-orbital mean-field ansatz could not reproduce the observed dynamics, since the 
build-up of a second and third mode is only possible by quantum correlations. Furthermore, although a 
TMBH description would 
give good approximative results, the dynamics within the ionic potential 
would be completely neglected. As a result, the above discussed fast oscillations in the left and right 
populations of the wells would not be traced in such a description. 

With this knowledge, we examined to which extent  the protocol proposed in Ref.~\cite{Gerritsma2012} to 
create an entangled state between the ion and the atoms works. We were able to 
show that this protocol still represents a viable strategy for the creation of such an entangled state. Due 
to the relatively short tunneling time, quantum correlations are not able to drastically harm the dynamics, 
and therefore the protocol. 
In particular, it turns out that in the STR a single mode description, that is, GP, works quite well.
In addition to this, we verified that the time-dependent dynamics of the protocol can be accurately well 
described by only two time-dependent mode functions, and therefore validating the theoretical approach used in 
Ref.~\cite{Gerritsma2012}.
The specific realization of the protocol chosen in this work is by no means optimal, but, nevertheless, we 
reached a fidelity of more than $90\%$ and according to our considerations on the experimental 
implementation we can conclude that the control of the junction will be possible on 
experimental relevant time scales.

\section{Acknowledgements}
The authors thanks Sven Kr\"onke and Valentin Bolsinger for many clarifying discussions.
This work has been financially supported by the excellence cluster 'The Hamburg Centre for
Ultrafast Imaging - Structure, Dynamics and Control of Matter at the Atomic Scale' of
the Deutsche Forschungsgemeinschaft. R.G. gratefully acknowledges support by the EU via the ERC (Starting 
Grant 337638) and the DFG via SFB-TR/49.


\appendix

\section{Derivation of the Uhlmann fidelity}\label{app:uhlman}

Here we briefly explain how the Uhlmann fidelity can be evaluated by using, instead of the systems density 
matrix, the one-body reduced density matrix.
Starting from Eq.~\eqref{eq:fidelity}, where now $\hat\sigma(t)$ is the  one-body reduced density matrix of 
the system at time $t$ and $\hat\rho_\mt{G}$ is the one-body reduced density matrix of the target state,  we 
can use the property of the Uhlmann fidelity, namely, it is preserved under a unitary transformation $U$ of 
$\hat\rho_\mt{G}$ and 
$\hat\sigma$ (thus $F(t;\hat\sigma,\hat\rho_\mt{G}) = F(t;U\hat\sigma U^\dagger,U\hat\rho_\mt{G}U^\dagger)$):
\begin{equation}
 F(t;\hat\sigma,\hat\rho_\mt{G}) = \mt{Tr} \left[  \sqrt{\sqrt{U\hat\rho_\mt{G}U^\dagger} 
U\hat\sigma(t)U^\dagger \sqrt{U\hat\rho_\mt{G}U^\dagger}  }  \right].
\end{equation}
Now choosing $U$ as the transformation making $\hat\rho_\mt{G}$ diagonal, corresponding to the natural 
orbital representation, the inner square roots can easily be evaluated. With the natural occupations 
$\lambda_k^\mt{G}$, the fidelity than can be written as
\begin{equation}
 F(t;\sigma,\rho_\mt{G}) = \mt{Tr} \left[ \sqrt{W}  \right].
\end{equation}
with the matrix $W_{kq} = \sqrt{\lambda_{k\vphantom{q}}^\mt{G}} C_{kq}(t) \sqrt{\lambda_q^\mt{G}}$, where 
$C_{kq}(t)=U\hat\sigma(t) U^\dagger$ is $\hat\sigma$ in the representation of the natural orbitals of 
$\hat\rho_\mt{G}$. 
Again introducing a unitary matrix which now makes $W$ diagonal $W = U_W D_W U_W^\dagger$, we can write the 
root of $W$ as $\sqrt{W} = U_W \sqrt{D} U_W^\dagger $. Using the cyclic permutation of the trace, the  
fidelity turns out to 
be
\begin{equation}
 F(t) = \sum_j \sqrt{w_j}
\end{equation}
with $w_j$ being the eigenvalues of $W$.


\bibliography{library}


\end{document}